\documentstyle[eqsecnum,aps,epsf,amssymb]{revtex}

\def\a{\alpha}
\def\b{\beta}
\def\ba{\bar{\a}}
\def\g{\gamma}
\def\d{\delta}
\def\s{\sigma}
\def\D{\Delta}
\def\X{\mbox{X}}

\begin{document}
\draft

\twocolumn[\hsize\textwidth\columnwidth\hsize\csname 
@twocolumnfalse\endcsname

\title{Kinetics of Nanopore Transport}

\author{Tom Chou}

\address{DAMTP,  University of Cambridge, 
Cambridge CB3 9EW, ENGLAND}

\date{\today}
\maketitle

\begin{abstract}

A nonlinear kinetic exclusion model is used to study
osmosis and pressure driven flows through nearly single
file pores such as antibiotic channels, aquaporins,
zeolites and nanotubules.  Two possible maxima in the
steady state flux as a function of pore-solvent
affinity are found.  For small driving forces, the
linear macroscopic osmotic and hydraulic permeabilities
$P_{os}$ and $L_{p}$, are defined in terms of
microscopic kinetic parameters.  The dependences of the
flux on activation energies, pore length and radius,
and driving forces are explored and Arrhenius
temperature dependences derived.  Reasonable values for
the physical parameters used in the analyses yield
transport rates consistent with experimental
measurements.  Experimental consequences and
interpretations are examined, and a straightforward
extension to osmosis through disordered pores is given.
 
\end{abstract}

\vspace{5mm}

\hspace{5mm}{\bf NOMENCLATURE}

\begin{equation}
\begin{array}{llll} 
a & \mbox{effective molecular diameter} & \a & \mbox{left entrance rate}\\
n_{T} & \mbox{total particle number density} & \b & \mbox{right exit rate}\\ 
n_{s} & \mbox{solute number density} & \g &\mbox{left exit rate}\\
k_{B}T & \mbox{thermal energy} & \d & \mbox{left entrance rate}\\
\ell & \mbox{section length}  & \lambda_{MFP} & \mbox{particle mean free path} \\
r_{p} & \mbox{pore entrance radius} & p(q) & \mbox{internal right(left) hop rate} \\
v_{T} & \mbox{thermal velocity}  & \xi & (\a,\b,\g,\d,p,q) \\
A_{p} & \mbox{effective pore entrance area}  & \chi_{0}^{(L,R)} & \mbox{left(right) solvent mole fraction} \\
E_{\xi} & \mbox{effective activation energies} &  \chi_{s}^{(L,R)} & \mbox{left(right) solute mole fraction} \\
J_{i}(t) & \mbox{flux between section}\, i\, \mbox{and}\, i+1\,(1/t) \quad
&  \sigma_{i} &\mbox{site $i$ occupation (0,1)}\\
J_{i} & \mbox{time average flux}\,  (1/t)& \tau_{\xi} &\mbox{molecular transit times}\, (\xi^{-1}) \\
L=N\ell & \mbox{pore length} & \D\mu_{0} &\mbox{solvent chemical potential difference}\\
L_{p} & \mbox{hydraulic permeability}  & \D P & \mbox{hydrostatic pressure difference} \\
N & \mbox{number of pore sections} & \D \Pi &\mbox{osmotic pressure difference}\\ 
P_{os} & \mbox{osmotic permeability} & 
\Gamma &\mbox{acoustic attenuation coefficient}\\ 
\: &\: & \X &\mbox{dimensionless driving} \\
\end{array}
\end{equation}

]

\section{Introduction}

Transport through microscopic pores is vital  to many biological and
industrial processes\cite{ALBERTS,FINK,BARRER}.  One phenomenon
involving flow through microscopic channels is osmosis. Ubiquitous
in biological functions such as cellular volume control, osmosis
and reverse osmosis are also exploited in industrial processes such
as solution concentration, filtration, and catalysis.

Living cells must transport nutrients, ions and water across their lipid
membranes.  Various biological structures and machinery, such as active
transporters, cotransporters, antiporters, and simple channels or pores have
evolved to perform these tasks \cite{ALBERTS,FINK}. Of central importance in
living organisms is the flow of water.  Certain cells, such as nephrons,
have a permeability to osmotically driven water transport too high to be
explained by simple permeation through an oily lipid membrane bilayer. In
addition to generic antibiotic pores such as gramicidin, pores which are
apparently water specific (aquaporin-CHIP) have been identified
\cite{AQUAREV}.  By inserting aquaporins into their membranes, cells can in
principle control overall osmotic permeabilities.  

Mechanisms of fluid transport in confined geometries are also relevant in
industrial applications, especially in separation processes. Both natural and
manmade materials such as zeolites contain many molecular-sized, essentially
single-file channels which can selectively absorb fluids. This size specificity
can be exploited in separation of a mixture of linear and branched chain alkanes,
where the zeolite acts like a sponge absorbing only the desired specie(s)
\cite{BARRER}. Confining particles in zeolite pores can also serve to catalyze
reactions \cite{KARGER,CUSSLER}. Therefore, there has been much research into
fluid structure and transport within confined pores
\cite{AUSTIN,DAVIS,MURAD,HAHN97,FISCHBARG95,FISCHBARG92,KOHLER,CRACKNELL,INDIAN,CHOUPRL}.

Although the equilibrium properties of osmotic ``pressure'' are understood in terms
of macroscopic forces \cite{GUELL} and statistical mechanics \cite{REICHL},
nonequilibrium osmotic transport has been less well categorized.  One approach
uses macroscopic flow equations (such as the parabolic Poiseuillean fluid velocity
profile through a pipe\cite{LANDAU}); these, applied to structures of molecular
dimensions lead to inaccurate prediction of membrane pore density, or pore radius,
particularly in biological examples \cite{FINK,GALEY}. Use of macroscopic
parameters such as viscosity, or hydraulic flow, to microscopic transport has led
to models that do not accurately describe most of the relevant microscopic
physical processes \cite{WEIN,WANG69,GALEY} and many conflicting viewpoints
\cite{FINK}.  A recent review of the historical controversies and misconceptions
is given by Guell and Brenner \cite{GUELL}, who present a clarifying macroscopic
description of osmosis.  Finkelstein \cite{FINK} also mentions the many conceptual
disagreements associated with osmosis and nonequilibrium thermodynamics,
particularly when applied to water transport across biomembranes.

In this paper, we consider molecularly-sized pore that are nearly  single-file. Many
natural examples of pores are of molecular sizes.  Zeolites typically have pores with radii
$\sim 2-10\mbox{\AA}$  and length from a few nanometers to millimeters.  Biological channels
Integral membrane proteins that form channels in biological systems are difficult to prepare for
X-ray crystallography, accurate dimensions of membrane water channels are not yet available.
However, spanning cell membranes, they are $\sim 50\mbox{\AA}$ in length. Based on sequential
analogies with antibiotic pores and electron cryocrystallography \cite{MITRA97,JAP} biopores
such as the water transporting aquaporins are believed to have pore diameters $\sim
2-4\mbox{\AA}$.  Thus, water movement inside these pores is statistically single-file, with
``overtaking'' rarely if at all occuring.

\begin{figure}[htb]
\begin{center}
\leavevmode
\epsfxsize=2.9in
\epsfbox{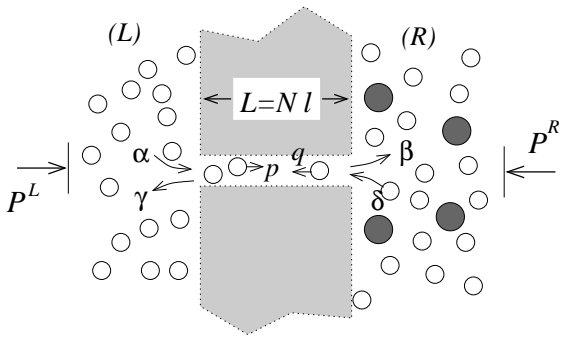}
\end{center}
\caption{Schematic of osmosis and pressure driven flow through membrane pores. The
reservoirs $(L)$ and $(R)$ are assumed infinite.  The rates
$\xi = (\alpha, \beta, \gamma, \delta, p, q)$ are conditional solvent entrance
probabilities at the ends of the pore indicated (see text).  For typical
experiments, the membrane is impermeable to solute, $\Delta
P = P^{R}-P^{L}$, $(L)$ contains pure solvent, 
and $J$ indicates the direction of solvent
flux.}
\label{Fig1}
\end{figure}

Using a symmetric exclusion model \cite{ASEM}, thermodynamics, and simple kinetic
theory, we formulate an approximate model for transport through nearly single-file pores
\cite{CHOUPRL} which can be solved exactly.  Estimates of  the microscopic kinetic
parameters used allow the theory to be general enough to predict trends and to explore
dependences on physical parameters such as pore geometry, microscopic interactions, and
bulk fluid properties. Our model complements, although is more flexible than detailed
numerical simulations.  Microscopic motions derived from simulations can in principle be
used to evaluate the system-dependent parameters used in this work, provided accurate
force fields are known.  

In the next Section, we review the usual linear phenomenological
expressions used to describe osmosis or pressure driven flows through
semipermeable membranes and identify all macroscopic variables. A one-dimensional
symmetric exclusion model is then formulated for particle dynamics within a pore
that connects two fluid reservoirs. The rate coefficients used are approximated
using kinetic theory near thermodynamic equilibrium.  

In the Results and Discussion, we explore the model as a function of reasonable
parameters and present the mean flow rates in a series of plots. The macroscopic
variables typically used are then related to the rate constants used in the
microscopic theory. Nonlinearities and Arrenhius temperature dependences are also
discussed.  In the Conclusions, we summarize our results and introduce extensions to
be studied in the following paper.  The Appendix gives a simple result for pores with
disordered interiors, or hopping rates.  

\section{Membrane and Pore Transport Models}

\subsection{Macroscopic Phenomenological Expressions}

Here, we briefly review the macroscopic descriptions which hitherto have
typically been used to model osmosis.  The typical starting
assumption is that the solvent flux is a function of the solvent
chemical potential difference between two reservoirs (see Figure 1)
that do not exchange solute. For two reservoirs with solvent at 
different {\it equilibrium} chemical potentials, 

\begin{equation}
J = J(\Delta\mu) =
\sum_{k=1}c_{k}(\Delta\mu_{0})^{k} = \sum_{k=1}d_{k}(-1)^{k}(\Delta \chi_{0})^{k},
\label{VIRIAL}
\end{equation}

\noindent where the last equality assumes $T^{L}=T^{R},\, P^{L}=P^{R}$,
and $c_k$ and $d_{k}$ are expansion coefficients which may depend on
external parameters such as temperature, pore-particle interactions, etc. They
may also depend on the {\it total} solute concentration in the reservoirs. 
Equation (\ref{VIRIAL}) assumes linear response and expands $\Delta\mu$ in
powers of {\it solvent} density (a virial expansion). 
The microscopic interactions are nearly identical for solvent in either reservoir
(nearly all virial terms for the solvent chemical potential cancel) except for the
infrequent encounters with solute in the solution $(R)$ side. To lowest order, the
(low){\it solute} concentration difference gives rise to the free energy difference
which can be seen by using $\chi_{s} = 1-\chi_{0}$, such that $J$ can be written in
terms of solute mole fraction difference $\Delta \chi_{s}\equiv
\chi_{s}^{R}-\chi_{s}^{L}$.  A similar expansion in hydrostatic pressure gives van't
Hoff's result that the solvent flow induced by an impermeable solute is
thermodynamically balanced ($J=0$) by a hydrostatic pressure difference equal to
$\Delta p = -\Delta \Pi= -k_{B}T(n_{T}^{R}\chi_{s}^{R}-n_{T}^{L}\chi_{s}^{L})
\simeq -k_{B}Tn_{T}\D\chi_{s}$, where $n_{T}\simeq n_{T}^{L} \simeq n_{T}^{R}$
are the total particle number densities.  Small currents between $(R)$ and $(L)$ are
thus assumed (fairly accurately) to be linear $J \approx L_{p}(\Delta\Pi- \Delta
p)$.  Under isobaric conditions, and $\chi_{s} \ll 1$, osmotic flow is often
represented by the osmotic permeability $P_{os}$ (also denoted $P_{f}$) where
$J\simeq P_{os}(n_{s}^R-n_{s}^L)$. 

The constants of proportionality, $L_{p}$ and $P_{os}$
determine the rates of flow and have been the subject of
considerable attention. Measurements of membrane permeabilities
and the number of pores per area, can yield single pore
conductivities.  Conversely, the number of pores in a cell or
vesicle can also be determined if single pore conductances can
be accurately measured or modelled.  Fully microscopic
approaches, such as molecular dynamics (MD) simulations reveal
detailed microscopic information on how fluids particles adsorb
to the pore interior\cite{SOKOL92,GUBBINS}; however, the time
scales readily achievable by MD ($\sim$ ns) cannot readily
measure steady state flows through a long pore, or allow
exploration of large parameter dependent flow regimes.  To
achieve channel flows in MD simulations, an external potential
(such as gravity or external field) is often applied to every
particle \cite{FLOWSHIT}, to approximate a convective velocity.
However, particles do not experience local ``pondermotive''
forces (as is the case in osmosis or pressure driven flows) in
the absence of external electric or gravitational fields.  In
confined nearly 1D systems, such external forces may yield
qualitatively different behavior (such as shocks in the
particle density \cite{ASEM,DERRIDA93}) from that expected in
hydraulic or osmotic flow.  

In the following, we analytically model particle dynamics through a single pore with the
assumption that the reservoirs are in thermodynamic equilibrium, as was assumed in the expansion
of $J$ (Eqn.  (\ref{VIRIAL})) about a thermodynamically defined $\D \mu_{0}$. We include
sufficient microscopic detail within the pore, while carefully obeying all thermodynamic
constraints in the connected reservoirs.

\subsection{One-dimensional Pore Exclusion Model}

Our single-file model is similar to but more general than
the one vacancy models of Hernandez \cite{FISCHBARG92} and Kohler
\cite{KOHLER}.  The approach is fully microscopic, and does not rely on
undefined hydrodynamic parameters (on these size scales) such  as drag
factors, viscosity, etc.  Nonetheless, results from the microscopic theory can 
be associated and directly compared with those from experimental
measurements.  Consider the pore shown in Fig.
\ref{Fig1} which depicts a  single-file or nearly single-file
pore. The main interactions that we include are the excluded particle
(solvent) volumes within the pore; thus we adapt one-dimensional
exclusion models \cite{ASEM,CHOUPRL}. The pore is divided into $N$ sections
labelled $i$, each of length $\ell$. The occupation variable $\s_{i}(t)$
defines whether  pore section $i$ at time $t$ contains a particle
($\s_{i}(t) = 1$) or is empty ($\s_{i}(t) = 0$).  

The discrete dynamics are defined by the follow rules: During any infinitesimal
time interval  $dt$, randomly chose a particle with label $0\leq i \leq N+1$,
where $i=0, N+1$ corresponds to choosing the $(L), (R)$ reservoir. If $i=0$ (the
left reservoir) is chosen during $dt$, {\it and} $\s_{1}(t) = 0$, then we fill
the first site $(\s_{1}=0) \rightarrow (\s_{1}(t+dt)=1)$ with probability $\a
dt$. If $i=N+1$  (the right reservoir)  {\it and} $\s_{N}(t) = 0$, then we
extract a particle from $(R)$ and set $\s_{N}(t+dt)=1$ with probability $\d dt$.
These steps allow injection of the end sites of the pore from the reservoirs if
and only if the end sites are empty at time $t$.  If a particle at $i=1$ is
chosen, we empty it into the $(L)$ reservoir with probability $\g dt$, and {\it
if} site $i=2$ is empty, we move the particle to the right with probability $p
dt$.  Similarly, if a particle at site $i=N$ is chosen, it moves into the $(R)$
reservoir with probability $\b dt$, and {\it if} $\s_{N-1}(t) = 0$, it moves to
the left with probability $q dt$. If an interior site, $1< i < N$, is chosen,
and  $i+1\, (i-1)$ is empty, then we move the particle to the right(left) with
probability $pdt  \,(qdt)$ . The model assumes that $p$ and $q$ are constant
along all the interior sites: Effects of site dependent $p=q$ are derived in
Appendix A.  All the moves are allowed only if the site to be entered is empty,
except for the reservoirs $i=0, N+1$. After the a step is chosen and particles
moved with the assigned probabilities, another particle is picked at random and
the possible moves made. This algorithm is identical to a Monte Carlo simulation
\cite{ALLEN} except that spatial moves are coarse-grained to $\ell$ and the
effective interaction energies (which determine the acceptance of moves) between
particles are infinite (representing hard spheres)  for a particle occupying the
neighboring site, or zero if the site is empty. These kinetic steps can be
summarized by the following instantaneous fluxes

\begin{eqnarray*}
J_{i}(t) = p\s_{i}(t)(1-\s_{i+1}(t))
-q\s_{i+1}(t)(1-\s_{i}(t))\hspace{0.3in} \\
\hspace{2.3in}1 < i < N-1
\end{eqnarray*}
\begin{equation}
\begin{array}{lr}
J_{(L)\rightarrow 1}(t) \equiv J_{0}(t) =  
\a(1-\s_{1}(t))-\g\s_{1}(t) & i=0 \\[13pt]
J_{N \rightarrow (R)}(t) \equiv 
J_{N}(t) = \b\s_{N}(t) - \d(1-\s_{N}(t)) &
i=N 
\label{RULES} 
\end{array}
\end{equation}

\noindent These dynamics implicitly include effective excluded volume
interactions between solvent particles, which are strong in  confined geometries
such as 1D pores. However, we assume that solvent-solvent attractions within the
pore are weak compared to solvent-pore particle attractions such that
concerted cluster hops are rare.  These processes can be be treated numerically
\cite{SHOLL} and is outside the scope of the present study.  

To extract the steady-state currents and occupations, we take the
time average $\langle \ldots \rangle$ of (\ref{RULES}) to find

\begin{equation}
\begin{array}{l}
\langle J_{i}(t)\rangle = p\langle\s_{i}(t)\rangle -q\langle\s_{i+1}(t)\rangle
+(q-p)\langle\s_{i}(t)\s_{i+1}(t)\rangle \\[13pt]
\langle J_{0}(t)\rangle = \a(1-\langle\s_{1}(t)\rangle) -
\g\langle\s_{1}(t)\rangle  \\[13pt]
\langle J_{N}(t)\rangle = \b\langle\s_{N}(t)\rangle -
\d(1-\langle\s_{N}(t)\rangle)
\label{JAVE}
\end{array}
\end{equation}

Exact solutions to the steady state $\langle J(t) \rangle$
under certain parameter regimes and in the thermodynamic limit
have been found using matrix algebra techniques
\cite{ASEM,DERRIDA93}.  However, in osmosis and hydraulic
pressure driven flows across symmetric pores, where the
particles do not experience intrinsic or pondermotive forces,
and have not developed collective stream velocities ({\it c.f.}
Conclusions), $q=p$.  That is, a particle is as likely to hop
to the left or right {\it if} both sites to the left and right
are unoccupied.  The only ``force'' driving osmosis is the
directionally asymmetric excluded volume along the length of
the pore, which is ultimately determined by the entrance and
exit rates at the boundaries $i=1, N$. Geometrically asymmetric
pores would have $p\neq q$ and need to be treated numerically.
The convective stream flow limit where also $p\neq q$, is
discussed in the Conclusions.  Many molecular dynamics
simulations designed to study flow in narrow channels impose a
microscopic pondermotive force with $p\neq q$ \cite{FLOWSHIT}
(via a gravitational potential for example) in order to
accelerate the flow.  However, gravity and pressure forces,
although macroscopically both derived from potentials, do not
manifest themselves microscopically in the same way. When $q=p$
the higher order correlation
$\langle\s_{i}(t)\s_{i+1}(t)\rangle$ vanishes in (\ref{JAVE})
and upon time averaging and invoking steady state particle
conservation, $J_{i} = J_{i+1} = J_{i+2} = \ldots = J_{0} =
J_{N}$, where we have dropped the $\langle\,\ldots \rangle$
notation. Since all the steady state currents across each
interior section is identical, we can sum them up along the
chain to obtain

\begin{equation}
J = {p \over N-1}(\s_{1}-\s_{N}) = J_{0} = J_{N}
\label{JJJ}
\end{equation}

Equations (\ref{JJJ}) give three equations for the three unknowns 
$J,\, \s_{1}$, and $\s_{N}$ which are found to be
(with $\xi \equiv (\a,\b,\g,\d,p)$)

\begin{equation}
J(\xi, N) = {p(\a\b-\g\d)\over (N-1)(\a+\g)(\b+\d) + p(\a+\b+\g+\d)}
\label{J}
\end{equation}

\noindent and

\begin{equation}
\s_{1} = {\a - J(\xi, N) \over \a+\g},\quad \s_{N} = 
{\d+ J(\xi, N) \over \b+\d},
\label{SIGMA}
\end{equation}

\noindent which determines $\s_{i}$,

\begin{equation}
\begin{array}{l}
\displaystyle \s_{i} = \s_{1} - (i-1){ J \over p}
= {\a \over \a+\b} \\[13pt]
\displaystyle\quad -{p(\a\b-\g\d) \over
(\a+\b)\left[(N-1)(\a+\g)(\b+\d)+p(\a+\b+\g+\d)\right]} \\[13pt]
\displaystyle\quad\quad\quad - {(i-1)(\a\b-\g\d)
\over (N-1)(\a+\g)(\b+\d) +p(\a+\b+\g+\d)}.
\end{array}
\end{equation}

\noindent For an infinitely thin membrane ($N=0$), the particles pass the
mathematical surface based upon their reservoir kinematics alone and never interact
with each other or the membrane interior while crossing the barrier.  Overtaking of
particles,  when two adjacent, indistinguishable particles switch positions, does not
contribute to the overall current. Therefore, the model above is also valid as long
as each section $i$ is not too wide as to contain more than one particle at any time.
Pore diameters of up to $2\sim 3$ times the particle diameters will on average
preclude multiple occupancy.

\subsection{Parameter Relationships} 

The parameters $\xi \equiv  (\a,\b,\g,\d,p)$ used in the above kinetic model are
conditional probabilities and reveal the structure and microscopic mechanisms of
osmosis and pressure driven flows. Here, we explicitly relate $\xi$ to macroscopic
thermodynamic variables.  Rather than find precise numerical values for $\xi$, we
estimate them using reasonable physical models and extract their most sensitive
dependences, thereby illustrating the qualitative features of osmosis and pressure
driven flows through a pore.  

The main assumption we make is that the entire system
is under local thermodynamic equilibrium (LTE):
Locally, particle velocities obey a Maxwellian
distribution.  This is justified since measurements of
almost all microscopic systems give single pore osmotic
transport rates of $J < 10^{7}/$s.  Typical pore
diameters in aqueous solutions give mean free paths
$\lambda_{MFP} \lesssim 5\mbox{\AA}$.  Thus, ambient
thermal velocities $v_{T} \simeq \sqrt{k_{B}T/m} \simeq
4\times 10^{4}$cm/s (where $m$ is the particle mass),
give mean collision times of $\tau_{coll} \simeq
\lambda_{MFP}/v_{T} \simeq 0.5\mbox{ps} \ll J^{-1}$. 
Therefore, particles within the pore equilibrate by
suffering $O(10^{5})$ collisions while being
transported from $(L)$ to $(R)$.  This is sufficient
for local thermodynamic equilibrium (LTE) to hold.

\begin{figure}[htb]
\begin{center}
\leavevmode
\epsfxsize=2.4in
\epsfbox{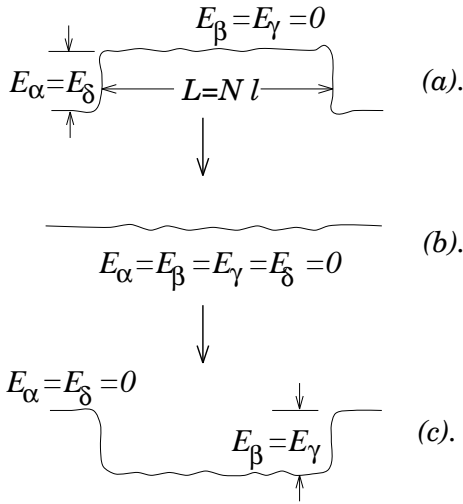}
\end{center}
\caption{Simplified pore-solvent interaction energy landscape for 
(a) repulsive and (b) attractive pores. The progression of the
activation energies $E_{\xi}$ are indicated as pores are altered or conditions modified
such that pore-solvent binding increases: $(a)\rightarrow(b)\rightarrow(c)$.}
\label{ENERGY}
\end{figure}

Figures \ref{Fig1}(b) are schematics  depicting  the activation energies $E_{\xi}$
determined by microscopic membrane-solvent interactions.  Although these are
predominantly of an enthalpic nature, determined by instantaneous intermolecular
potentials, a small entropic contribution may arise due to for example rotational
averaging of nonspherically symmetric particles. We do not attempt to model in detail
all the configurational dependences of particle trajectories, but rather assume the
activation energies to be effective quantities, averaged over say, molecular rotations,
internal vibrations, etc, which will not qualitatively affect the mean quantities we
wish to study. $E_{p, q}$ are internal binding energies to the pore walls which must be
overcome when particles hop from one section to another and can be thought of as
effective interaction energies between pore and solvent. As mentioned, the model
presented is valid provided solvent-solvent attraction within the pore $\ll E_{p}$.
$E_{\b, \g}$ and $E_{\a, \d}$ are the effective energies of removing and injecting a
particle. If we assume that there is no additional ``constriction'' at the pore ends,
for a pore that repels particles {\it relative} to its interaction energies in the bulk
reservoirs $(L)$ and $(R)$, $E_{\b, \g} \simeq 0$ and $E_{\a, \d} > 0$ (Fig
\ref{ENERGY}(a)).  Similarly, as shown in Fig \ref{ENERGY}(b), an attractive pore will
have $E_{\b, \g} > 0$ and $E_{\a, \d} \simeq 0$. The physical origin of the energies
such as $E_{\alpha}-E_{\gamma}$ defined in Figure 2(b) are most likely from hydrogen
bonding interactions in aqueous systems.  Transferring a water molecule from the polar,
hydrogen bonding bulk environment into the pore region would require breaking hydrogen
bonds ($\sim 5k_{B}T $ each) and forming van der Waals-type interactions with the pore
interior.  

{\it Additional} activation energies at the
entrance/exit sites would represent for example
energetics of a possible intermediate state where
hydrogen bonds are being broken, but attractive van der
Waals interactions have not yet formed. For simplicity
of notation, we assume these are small and neglect
them.  Inclusion of these boundary activation barriers
will only shift the absolute values of $E_{\a}, E_{\b},
E_{\g}, E_{\d}$ and will not qualitatively change our
results.
 
The probabilities $\xi$ can be estimated by first passage time or transition state
theory. The average time $\tau_{\xi}$ for a thermally agitated particle to climb over
a barrier of height $E_{\xi}$ is given by

\begin{equation}
\tau_{\xi} \simeq \left[\xi_{0}\exp(-E_{\xi}/k_{B}T)\right]^{-1}
\label{TAUXI}
\end{equation}
 
\noindent where $\xi_{0}$ is a thermal attempt frequency.  The transition
probabilities per time $\xi dt$ are thus given by $dt/\tau_{\xi}$. Upon choosing $dt
\ll \tau_{\xi}$,

\begin{equation}
\xi \simeq \xi_0 \exp(-E_{\xi}/k_{B}T).
\end{equation}

\noindent The prefactors of $q=p$ and $\b, \g$, are estimated by $p_{0}, \b_{0},
\g_{0} \simeq v_{T}/\ell$. Although more sophisticated transition state theories
exist, the thermal attempt frequencies are all on the order of  1ps$^{-1}$. The
frequencies $\a_{0}$ and $\d_{0}$ however, depend on the thermodynamic state of, and
the particle numbers in the reservoirs $(L)$ and $(R)$ respectively, as well as the
pore entrance areas. For example, for effusing dense gases, $\a_{0} = n_{T}^{L}
v_{T}^{L} A_{p}^{L}/4$, proportional to the number density $n_{T}^{L}$ of particles
able to enter, the thermal velocity $v^{L}_{T}$, and the available area $A^{L}_{p}$
for entrance.  A completely analogous expression holds for $\d_{0}$ with $(L)$
replaced by $(R)$.  Although the $1/4$ factor is valid for gases, a qualitatively
similar expression holds for dense liquids.  

The microscopic pore area $A_{p}$ can be defined by
$E_{\a, \d}$. First consider $E_{\a}(\vec{r})$ to be a
general interaction of pushing a particle
perpendicularly through a membrane. When the particle
is pushed through at a position $\vec{r}$ away from the
pore (through a lipid layer for example),
$E_{\a}(\vec{r}) \simeq \infty$.  When the trajectory
runs through the center of an approximately circular
pore entrance, $E_{\a} \equiv E_{\a}(\vec{r}=0)$. 
Therefore, we can define a pore radius $(A_{p} \simeq
\pi r_{p}^2)$ by demanding activation energies
such that trajectories outside of $r_{p}$ are
energetically rare. For example,

\begin{equation}
E_{\a}(r < r_{p}) - E_{\a}(0) < \, \sim\! 10k_{B}T.
\label{AREA}
\end{equation}

\noindent Condition (\ref{AREA}) simply contrains $r<r_{p}$ to be within a region
where entrance is not impeded by ``wall'' interactions. Thus, $E_{\a}$ (and $E_{\d}$)
are the activation energies for a particle dragged through the approximate
``centerline'' of the pore. When more than one species occupies the reservoirs, but
only one (the solvent) can enter the pore, the entrance rates will also be
proportional to their mole fractions. For notational simplicity, we henceforth define
$\a \equiv \a_{0}\chi_{0}^{L}\exp[-E_{\a}/k_{B}T]$ and $\d \equiv
\d_{0}\chi_{0}^{R}\exp[-E_{\d}/k_{B}T]$. Thus, $\a_{0}\propto A_{p}$ is now the rate
prefactor for pure solvent entrance ($\chi_{0} = 1$) at a reference number density
for a pore with entrance area $A_{p}$.  Only the solvent fraction can enter the pore
and partake in the dynamics described by(\ref{RULES}) because the solutes are too
large to enter through $A_{p}$.  

Although $q=p$, a net current will occur due to  differences in $\a \neq \d$, and/or
$\b \neq \g$. Henceforth, for simplicity, we consider only structurally right
cylindrical pores such that $\b_{0} = \g_{0}$ and $\a_{0}=\d_{0}$ (or $A_{p}^{L} =
A_{p}^{R}$ and $E_{\a}=E_{\d}$) under equilibrium ($T^{L}=T^{R}$ and $P^{L}=P^{R}$)
conditions. Define from the numerator of $J(\xi, N)$ the dimensionless driving
variable 

\begin{equation}
\begin{array}{ll}
\displaystyle \X & \displaystyle \equiv {\a\b-\g\d \over 
\a\b } \\[13pt]
\displaystyle \: & = \displaystyle
1 - {\chi_{0}^{R}\over \chi_{0}^{L}}\exp\left[
(E_{\a}+E_{\b}-E_{\g}-E_{\d})/k_{B}T\right] \\[13pt]
\displaystyle \: & \displaystyle
= 1-e^{\D E/k_{B}T} + {\D \chi_{s} \over 1-\chi_{s}^{L}}e^{\D E/k_{B}T}
\label{DELTA}
\end{array}
\end{equation}

\noindent where $\D E \equiv E_{\a}-E_{\g}+E_{\b}-E_{\d} \equiv E^{R} - E^{L}$,
$\D \chi_{s} \equiv \chi_{s}^{R}-\chi_{s}^L$, and $\chi_{s}^{(R), (L)}$ is the
solute mole fractions in $(R), (L)$. The solute particles, unlike the solvents,
are too large to enter the pore.  There will be a steady state particle flux as
long as $\X \neq 0$, which can occur either by virtue of  $\D E \neq 0$
(predominately pressure driven flow), and/or $\D \chi_{s} \equiv
\chi_{s}^{R}-\chi_{s}^{L} \neq 0$ (osmosis).  

%Since the function $J(\xi, N)$
%has nonlinear $\a,\b,\g,\d$ dependences in the denominator, it is a 
%nonlinear function of $\X$, $\chi_{s}$, and $E_{\xi}$ (Appendix A).  

Although electrostatic potential differences can contribute to $\D E_{\xi}$,
they would also make $q\neq p$, which we do not consider. Therefore, we assume
that $\D E_{\xi}$ arise solely to a hydrostatic pressure difference in the baths
which pushes the molecules against their repulsive interactions, and change the
relative energies of activated pore entrance.

%Taylor expansion of $\X$ about $(\D E = \D P = 0)$
%yields

%\begin{equation}
%\D  \simeq {\D \chi_{s}\over 1-\chi_{s}^{L}}
%- {1-\chi_{s}^{R}\over 1-\chi_{s}^{L}}\left({\partial \D E\over
%\partial \D P}\right)_{\D P = 0}{ \D P\over k_{B}T} + O(\D P)^2
%\end{equation}

\section{Results and Discussion}

In this section we  examine the solutions of one dimensional models and
discuss their physical meaning. Various effects of physical and chemical
parameters, such as pore length and solute interaction dependences, are
outlined. 

\subsection{Osmosis}

Consider symmetric pores ($\b = \g,\, E_{\a} = E_{\d}$) connecting two reservoirs under
identical hydrostatic pressures so that $\D E = 0$. For simplicity, we will also assume
pure solvent in $(L)$ such that  $\D \chi_{s} \equiv \chi_{s}^{R}$.  Even though particle
enthalpies are identical for $(L)$ and $(R)$,  $\a_{0} \neq \d_{0}$ will drive an
osmotic flow from $(L) \rightarrow (R)$ since in this limit $\X \equiv (\a-\d)/\a
\simeq \chi_{s}^{R}\neq 0$ is due solely to a solute concentration difference. The
steady state current becomes

\begin{equation}
J(\xi, N) \simeq {\ba p \X \over (N-1)(\ba+1)(\ba+1-\ba \X)
+\bar{p}(2\ba +2-\ba \X)},
\label{JBAR}
\end{equation}

\noindent where $\ba \equiv \a/\b$. Since $\a/\b$ is  a ratio of pore entrance to
exit rates across the pore, it measures an effective pore-solvent binding constant.  

How does $J(\xi, N)$ behave as pore-solvent combinations with varying $\ba$ are
used? When a symmetric pore is repelling, the energy landscape is approximately that
shown in Fig. \ref{ENERGY}(a), where $E_{\a} \simeq E_{\d} > 0$, and $E_{\b} \simeq
E_{\g} \simeq 0$.  As pores with increasing solvent  affinities are used,
$E_{\a}\simeq E_{\d}$ decrease while  $\a,\, \d$ increase, until the energy barrier
shown in Fig.  \ref{ENERGY}(b). is reached, where $\ba = \a_{0}/\b_{0}$. Further
increasing pore-solvent attraction requires $E_{\a}\simeq E_{\d}\simeq 0$, and
increasing $E_{\b}\simeq E_{\g}$.  Thus, as pores follow the sequence $(a)\rightarrow
(b) \rightarrow (c)$, $\ba$ first increases as $\exp(-E_{\a}/k_{B}T)$ (as $E_{\a} > 0$ is
decreased), until $\ba = \a_{0}/\b_{0}$, where upon $\ba$ increases according to
$\exp(E_{\b}/k_{B}T)$ (as $E_{\b}>0$ is increased). While the pore is repelling, and
$\b=\b_{0}$ is constant, $J(\xi, N)$ has a maximum as a function of pore-solvent
affinity at 

\begin{equation}
\ba^{*}(\b_{0}) \equiv \left({\a^{*}\over \b_{0}}\right)
=\left[{2p/\b_{0}+(N-1) \over (N-1)(1-\X)}\right]^{1/2}.
\label{MAXIMUMA}
\end{equation}

\noindent However, if $\a_{0} < \a^{*}(\b_{0})$, the maximum defined by
(\ref{MAXIMUMA}) is not reach by decreasing $E_{\a}$. Further increasing $\a/\b$
requires decreasing $\b < \b_{0}$ with fixed $\a = \a_{0}$ as shown by
the transition (b)$\rightarrow$(c) in Fig. \ref{ENERGY}.  The value of $\b$ which yields a
maximum in $J$ in this case is

\begin{equation}
\ba^{*}(\a_{0}) \equiv \left({\a_{0}\over \b^{*}}\right)
= \left[{\a_{0}(N-1) \over \a_{0}(N-1)(1-\X) + p(2-\X)}\right]^{1/2}
\label{MAXIMUMB}
\end{equation}

\begin{figure}[htb]
\begin{center}
\leavevmode
\epsfysize=4.2in
\epsfbox{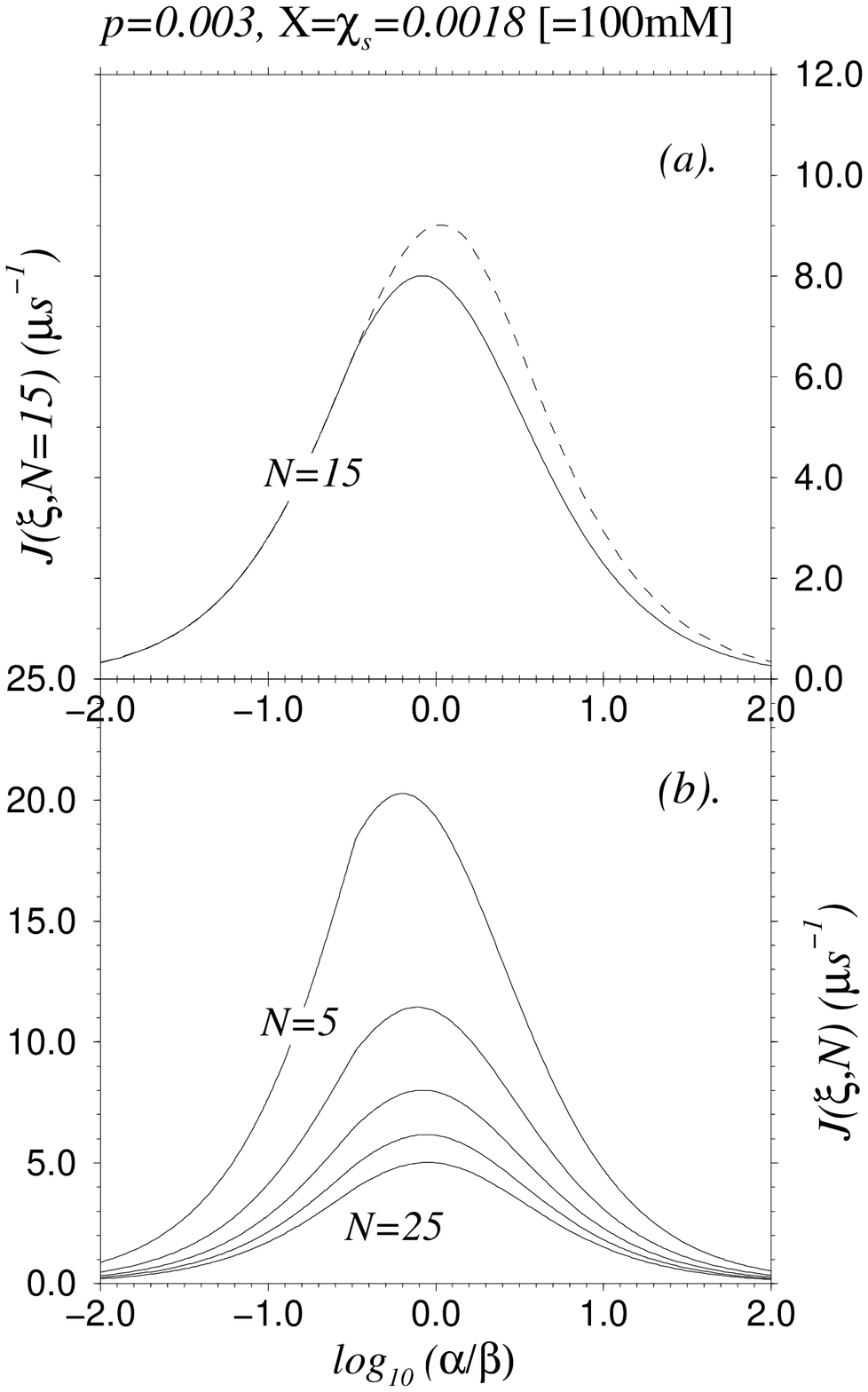}
\end{center}
\caption{Flux as a function of effective pore-solvent binding $\ba$. We use
representative values $dt\simeq 10$fs, $p=\b_{0}=0.003$, and $\X=\chi_{s}^{R}
\simeq 0.0018$, corresponding to osmosis from pure solvent to a 100mM solute solution.
(a). $J(\xi,N=15)$ is shown for $\a_{0}=0.001$(solid curve) and $\a_{0}=0.0045$(dashed
curve) which give difference maxima at $\ba^{*}(\a_{0})$ and $\ba^{*}(\b_{0})$
respectively. (b) shows $J$ for $N=5,10,15,20,25$ when $\a_{0} = 0.001$.
Note the discontinuities in slope at $\log (1/3)$ corresponding to the 
point where the source of increasing affinity switches 
from the increasing of $\a$ to
the decreasing of $\b$. (Figure \ref{ENERGY}(b).) }
\label{ALPHA}
\end{figure}

Figure \ref{ALPHA} shows the current  as the solvent-pore affinity is increased
according to $\ba$, as $\ba = \a_{0}/\b \rightarrow \a_{0}/\b_{0}(\equiv
\ba_{0})\rightarrow \a/\b_{0}$. The solute concentration is represented in molar units
$\left[c_{s}\right]= \chi_{s}^{R}\times 55.56$, such that a 100mM aqueous solute
solution corresponds to $\X = \chi_{s}^R\simeq 0.0018$.  Although $J$ and $\xi$ can
all be rescaled by the constant $p$, for concreteness and comparison with
experiments, we consider a time step $dt \simeq 10\mbox{fs} \ll \tau_{p}\approx
0.3$ps, which sets $\b_{0} \approx p \simeq 0.003$ and yields flows in the range  of
$10^{8}\times J$ (particles per $\mu$s) in good agreement with $P_{os} \sim
10^{-14}$cm$^{3}$/s measured across aquaporin water channels with physiological
($\sim 100$mM) osmolyte solutions \cite{FINK,AQUAREV}.

Figure \ref{ALPHA}(a) plots the flux ($\mu\mbox{s}^{-1}$) as a function of $\log \ba$
for two different values, $\a_{0}/\b_{0}=0.001/0.003 < (\a/\b_{0})^{*} =
\sqrt{8/7(1-\X)}$ and $\a_{0}/\b_{0} = 0.0045/0.003 > (\a/\b_{0})^{*}$. At $\log (1/3)$,
the curve bifurcates into the two solutions corresponding to either decreasing $\a$ or
increasing $\b$, with maxima determined by (\ref{MAXIMUMA}) and (\ref{MAXIMUMB})
respectively.  The discontinuity in slope at $\log(1/3)$ (dashed curve) is apparent,
although  the one at $\log(3/2)$ near the maximum is not. Figure \ref{ALPHA}(b) compares
$J(\xi, N)$ for various pore lengths $L=N\ell$ at $\a_{0} = 0.001$ (such that
(\ref{MAXIMUMB}) pertains). Whether $\a_{0}/\b_{0}$ is greater or less than
$\ba^{*}(\b_{0})$ depends on the reservoir number density and $A_{p}$; for gases,
$\a_{0}\ll \b_{0}$ and a curve with maximum defined by (\ref{MAXIMUMB}) obtains. 
However, dense liquids entering large pores may have relatively large entrance rates
such that $\a_{0}/\b_{0} \gtrsim \ba^{*}(\b_{0})$ and $J(\xi,N)$ may have a maximum
defined by (\ref{MAXIMUMA}).  From (\ref{MAXIMUMA}), this latter condition is less
likely for larger driving $\X$, and thus $\ba^{*}(\b_{0})$.  

\begin{figure}[htb]
\begin{center}
\leavevmode
\epsfxsize=2.8in
\epsfbox{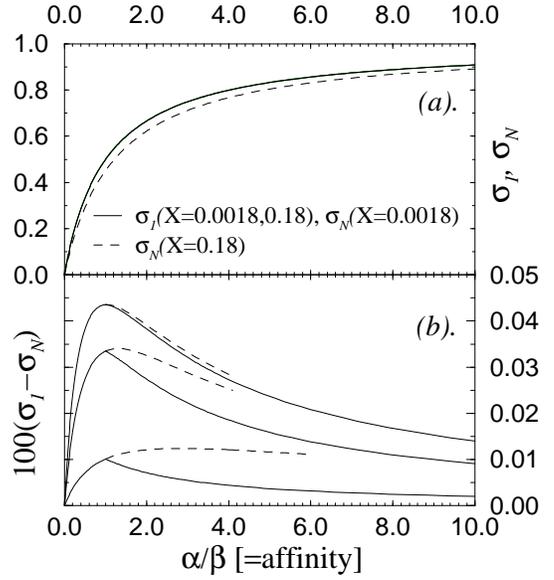}
\end{center}
\caption{Occupations $\sigma$ for $p=\b_{0}=\a_{0}=0.003$. 
(a) Pore occupations for $\X=0.0018$ are all indistinguishable 
on this scale (solid curve). Only at $\X=0.18$(10M solution) is 
$\sigma_{N=30}$ (dashed curve) sufficiently different from $\sigma_{1}$.
(b) 100$\times(\s_{i}-\s_{N})$ as a function of $\a/\b$ for $p=0.003, 0.03, 0.3$.
Note the different behavior when $\a=\a_{0}=0.003$ at $\a/\b = 1.0$. 
The dashed fragments represent the occupation differences if $\a_{0} > 0.003$. }
\label{TAU}
\end{figure}

Figure \ref{TAU}(a) shows the average occupation $\sigma_{i}(\ba)$ for $\X =
\chi_{s} = 0.0018$. The dependence on spatial position  $i$ is weak and is not apparent
on the scale of Fig. \ref{TAU}(a), except if $\X=0.18$ (corresponding to 10M solutions)
when $\s_{N}$ can be distinguished from $\s_{1}$.  Therefore, from Eqn.
(\ref{SIGMA}) $\a/(\a+\b)$ is an approximate value of averaged occupations at small
$\X$.  The small difference in occupation along the chain is consistent with the
assumption of local thermodynamic equilibrium.  Figure \ref{TAU} shows the
difference $10^{3}\times(\sigma_{1}(\ba)-\sigma_{N}(\ba))$ with  $\a_{0}/\b_{0} = 1.0$
(solid curves). The dashed curves represent the  occupation difference
$10^{3}\times(\s_1-\s_N)$ if $\a_{0}/\b_{0} \gg 1$.  

The behavior of $J(\xi, N)$ as a function of affinity $\ba$ can be further
understood in terms of the associated pore occupations. At low $\ba$, the pore is
repelling and contains few if any particles that can be transported. At high
affinities, and occupations, most internal hopping steps are not fulfilled due to
occupied neighboring sites choking off the flow. Only at intermediate $\ba$ and
near (\ref{MAXIMUMA}) or (\ref{MAXIMUMB}), and intermediate occupations  will
maximum flow occur. At larger $\X$, $\s_{N}$ decreases (see Fig. \ref{TAU}(a)),
decreasing overall pore occupation and increasing the critical affinities
$\ba^{*}(\b_{0})$ and $\ba^{*}(\a_{0})$ where the maximum occurs. Therefore, the
condition represented by (\ref{MAXIMUMB}) becomes more relevant for larger osmotic
pressures.  

\begin{figure}[htb]
\begin{center}
\leavevmode
\epsfxsize=2.9in
\epsfbox{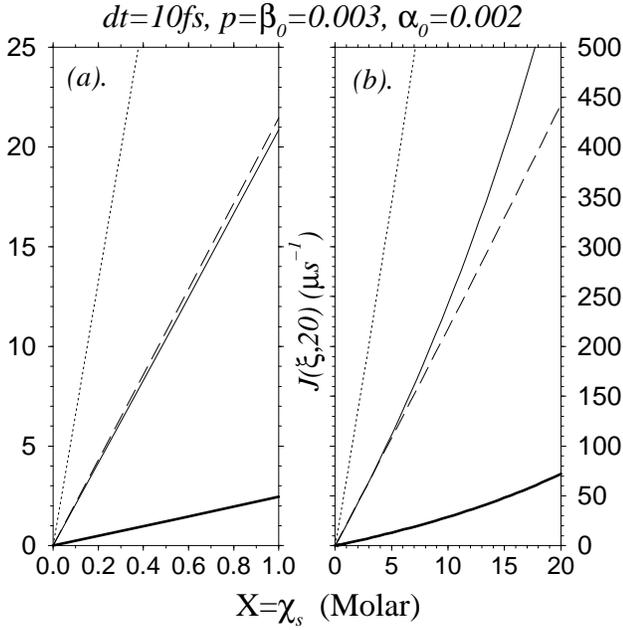}
\end{center}
\caption{The driving force dependence of $J(\xi,20)$ for osmosis from pure solvent to
solution with $p=\b_{0}=0.003$ and $\a_{0}=0.002$.  
Various values of $\ba=\a/\b$ are shown.
(a) For solute concentrations $\lesssim 5$M, the 
current is linear in $\X=\chi_{s}^{R}$. (b) At higher solute concentrations, 
nonlinearities in $J(\xi, 20)$ arise, particularly for high values of $\ba$
an pore occupation.}
\label{NONLINEAR}
\end{figure}

Figures \ref{NONLINEAR} show the particle flux per microsecond, $10^{8}\times J(\xi, N=20)$,
as a function of $\X \simeq \chi_{s}^{R}$ at $\b_{0}=p=0.003,\, \a_{0}=0.002$ for various
pore-solvent affinities $\ba$. For the low values $\ba=0.01$(dashed curve) and $\ba =
0.1$(dotted curve) the pore is repelling,  while for $\ba = 1.0$(solid curve) and
$\ba=100$(thick curve) $\a=\a_{0}$ while $\b$ decreases from $\b_{0}$. The solid curve with
$\ba =1.0$ is near the maximum defined by (\ref{MAXIMUMB}) while $\ba = 100$ is within the
choked flow limit represented by the decreasing branch on the high $\ba$  side of Figures
\ref{ALPHA}(a,b). The flow is linear for low $\X$ as shown in \ref{NONLINEAR}(a). Figure
\ref{NONLINEAR}(b) clearly shows that nonlinearities are important for large $\ba$ and that
the flow for $\ba = 1.0$(solid curve) is lower that that for $\ba = 0.01$(dashed curve) only
for small driving X. As the nonlinearity sets in for $\ba = 1.0$, the flux increases faster
than that for $\ba=0.01$ and surpasses it near X$\sim 3$.  The flow nonlinearities shown in
Fig \ref{NONLINEAR}(b) arise at large solute concentration differences and are only those 
resulting  from  the nonlinear dynamics determined by equations (\ref{RULES}) and do {\it
not} include solute nonidealities, unstirred layers, etc.  That is, the coefficients 

\begin{equation}
d_{k} = {\ba^{k}p\left[(\ba+1)(N-1)+\bar{p}\right]^{k-1}\over 
(\ba+1)^{k}\left[(\ba+1)(N-1)+2\bar{p}\right]^{k}}
\label{DK}
\end{equation}

\noindent contain only the kinetic parameters describing the {\it solvent}-pore interactions.
Furthermore, our treatment for $\a$ and $\b$ assume thermodynamic equilibrium, or no
unstirred/polarization layers.  The modifications necessary to include unstirred layers are
treated in the following paper.  

From the denominator of (\ref{JBAR}), $J$ is linear in $\X$ for $\ba \X/(\ba +1) \ll 1$.  From
(\ref{SIGMA}), we see that $\ba/(\ba+1)$ is the average occupation for small $J$. Therefore,
nonlinearities are important only when occupation $\sigma_{i}$ is high, when exclusion
interactions, and thus nonidealities, are most pronounced (the $\ba = 1.0, 100$ curves in Fig.
\ref{NONLINEAR}(b)).  Although Eqn. (\ref{JBAR}), or (\ref{DK}) fully describes particle
exclusion nonidealities within the pore, for the small $\D \chi_{s}$ usually encountered, a
linear relationship,

\begin{equation}
J(\xi, N)\simeq {\ba \over \ba+1}{p\D \chi_{s} \over (N-1)(\ba+1)+2\bar{p}}
\label{JLINEAR}
\end{equation}

\noindent is accurate as demonstrated by Fig \ref{NONLINEAR}(a).  In the limit $\bar{p}
\ll (\bar{\alpha}+1)N$, $P_{os}$ approaches 

\begin{equation}
P_{os} \cong \left({a\over L}\right){\alpha\bar{p} \over (\bar{\alpha}+1)^{2} 
n_{T}}.
\label{POS1}
\end{equation}

\noindent In this limit(\ref{POS1}),
the ``rate limiting steps'' are the internal steps $(p)$ of pore particles,  and the overall flux scales as $L^{-1}$.  In the limit represented by
(\ref{POS1}), an Arrhenius temperature dependence of
$(E_{\alpha}-E_{\beta}-E_{q})/k_{B}T \left[(E_{\beta}-E_{\alpha}-E_{q})/k_{B}T\right]$
is expected for $\ba\gg 1 \left[\ba \ll 1\right]$.  

According to the form of $\a_{0}$,  $P_{os} \propto \alpha\bar{p} \propto r_{p}^{2}$
for $\bar{\alpha}\ll 1$, and $P_{os} \propto p/\bar{\alpha} \propto r_{p}^{-2}$ for
$\bar{\alpha}\gg 1$.  Note the curious $P_{os} \propto r_{p}^{-2}$ dependence
indicating the $\bar{\a} > \bar{\a}^{*}$ regime shown by the large
$\log_{10}\bar{\a}$ regions of Fig. \ref{ALPHA} where decreasing $r_{p}$ (and $\a$)
increases $\bar{J}$ by decreasing the high pore occupation.

Now consider the limit $\bar{p} \gg (\bar{\alpha}+1)N$. The rate limiting steps now
involve pore entrance or exit and 

\begin{equation}
P_{os} \cong {\alpha \over 2(\bar{\alpha}+1)n_{T}},
\label{POS2}
\end{equation}

\noindent independent of $L$. This limit is equivalent to the single site $(N=1)$ model
since only the boundary entrance and exit rates are relevant. The radius and
temperature dependences in this case are $P_{os} \propto \alpha \propto
r_{p}^{2}\exp(-E_{\alpha}/k_{B}T)$ and $P_{os} \propto \beta \propto
\exp(-E_{\beta}/k_{B}T)$ for $\bar{\alpha}\ll 1$ and $\bar{\alpha}\gg 1$
respectively.  Thus, the pore length  $L\simeq N\ell$ can determine the temperature
dependence of $J$ since it determines which of the two limits (\ref{POS1})  or
(\ref{POS2}) is relevant. Except for experiments where lipids undergo phase
transitions which affect the pore structure \cite{BOEHLER}, a single exponential
temperature dependence is almost always experimentally observed in biological
systems involving osmosis \cite{COHEN}, indicating one of the above limits for
$P_{os}$ holds in the regimes explored. One interpretation is that a number of
hydrogen bonds ($\sim 5k_{B}T$ each) must be broken before water can enter the
pore. This is consistent with the identification of $E_{\alpha}$ as the energy
required to break water H-bonds in the $\bar{\alpha}\ll 1$ limit of (\ref{POS1}).
Fluxes of nonpolar solvents through artificial pores may be expected to display richer
temperature dependences since a wider variety of temperatures and molecular
interactions can be accessed and experimentally probed.

\subsection{Pressure Driven Flow}

\begin{figure}[htb]
\begin{center}
\leavevmode
\epsfxsize=2.4in
\epsfbox{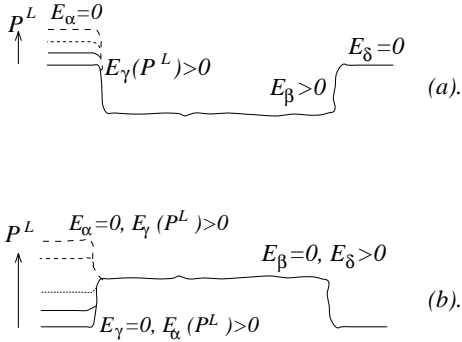}
\end{center}
\caption{Hydrostatic pressure increases in left reservoir. (a) 
In attractive pore, $\a=\d$, $E_{\b}$ is constant, and $E_{\g}=E_{\g}(P^L)$. (b)
For repelling pore, and small pressure differences (lower solid and dotted levels)
$\g=\b,\, E_{\d}$ is fixed, and $E_{\a}=E_{\a}(P^L)$. Pressure differences larger than 
$E_{\d}$ result in $E_{\a} \approx 0$ and $E_{\g}=E_{\g}(P^L)>0$.}
\label{ENERGYP}
\end{figure}

For incompressible reservoirs without solutes, $\D
\chi_{s} = \chi_{s}^{L} =\chi_{s}^{R} = 0$; however,
flow can be driven by differences in hydrostatic
pressure.  Hydrostatic pressure variations, rather than
controlling the rates via the number density in the
pre-exponential factors $\a_{0},\, \d_{0}$, modify the
relative activation energies $E_{\xi}$. For example,
consider increasing $P^{L}$ keeping $P^{R}$ constant.
For nearly incompressible liquids, the particle density
changes negligibly, but their particle enthalpies
increase as they are pushed against their interaction
potentials. Changing the relative reservoir and pore
enthalpies changes the entrance and exit kinetic
coefficients according to Fig. \ref{ENERGYP}(a), (b)
for initially repelling or attractive pores.

\noindent Since we assume only $P^{L}$ is varied, $E_{\d},\, E_{\b}$ are kept constant,
while $E_{\a} = E_{\a}(P^{L})$ and $E_{\g} =E_{\g}(P^{L})$.  First consider an
attractive pore where only $E_{\g}(P^{L})$ and $\g$ changes as $P^{L}$ is increased, as
schematically depicted in Fig.  \ref{ENERGYP}(a). The flux is

%\begin{figure}
%\begin{center}
%\leavevmode
%\epsfxsize=2.3in
%\epsfbox{ENERGYP.eps}
%\end{center}
%\caption{Energy barriers as pressure $P^{L}$ is increased. 
%(a). A repelling pore, where $E_{\a}(P^{L}),\, E_{\g}(P^{L})$
%depend on the magnitude of $P^{L}$. (b). An attractive pore where 
%$E_{\a} = E_{\d} \simeq 0,\, E_{\b}>0$, and only $E_{\g}$ 
%increases with $P^{L}$. }
%\label{ENERGYP}
%\end{figure}

\begin{equation}
\begin{array}{l}
\displaystyle J\left(\g(P^{L}),N\right) = \\
\displaystyle \quad \quad
{p\ba \X \over (N-1)(\ba+1-\X)(\ba+1)
+\bar{p}(2\ba + 2- \X)}
\label{ATTRACTP}
\end{array}
\end{equation}

\noindent where

\begin{equation}
\X(P^{L}) \equiv 1-\exp\left[\D E/k_{B}T\right]
\end{equation}

\noindent and $\D E \equiv E_{\b}-E_{\g}(P^L)$. 

When hydraulic pressure pushes particles through a repelling pore, the current can be
written in the form

\begin{equation}
\begin{array}{l}
\displaystyle J\left(\a(P^L), \g(P^{L}),N\right) = \\
\displaystyle {p\bar{\d}\X \over (N-1)(\bar{\d}+1)(\bar{\d}+1-\X)
+\bar{p}\left(\bar{\g}+\bar{\d}+(\bar{\d}+1)(1-\X)/\bar{\g}\right)} 
\label{REPELP}
\end{array}
\end{equation}

\noindent where in this case, 

\begin{equation}
\X = 1-\exp\left[(E_{\a}(P^L)+E_{\b}-E_{\g}(P^L)
-E_{\d})/k_{B}T\right],
\end{equation}

\noindent and $\bar{d}$ measures the solvent-pore affinity since $\ba$
is varying with pressure.  For low pressure differences such that 

\begin{equation}
P^L - P^R \ll E_{\d}
\left[\left({\partial \D E \over \partial 
\D P}\right)_{T, \D P=0}\right]^{-1}
\end{equation}

\noindent flow across repelling pores can be further simplified since $\g = \b$ as
evident in the lower three pressures $P^L$ in Fig. \ref{ENERGYP}(b).  When the pressure
pushes the enthalpy of the left bath above that of the pore, the full pressure
dependence (\ref{REPELP}) is required.  Equations (\ref{ATTRACTP}) and (\ref{REPELP}),
along with the equation of state of the solvent in the reservoirs completely determine
the nonlinear flux-pressure relationship. However, for low pressures, a Taylor
expansion about $P^L=P^R$ yields 

\begin{equation}
\X \approx -{\D P\over k_{B}T}\left({\partial \D E
\over \partial \D P}\right)_{T,\D P=0} = -\tilde{v}{\D P\over k_{B}T},
\end{equation}

\noindent where the last equality is valid for incompressible liquids and $\tilde{v}$
is the molecular volume of a solvent particle in the reservoirs (for H$_{2}$O,
$\tilde{v} \simeq 3\times 10^{-23}\mbox{cm}^{3}$). Linearizing (\ref{ATTRACTP}) and
(\ref{REPELP}) about small pressure differences, and recalling that $\a \simeq \d$, we
find 

\begin{equation}
L_{p} \cong {P_{os} \over k_{B}T}
\end{equation}

\noindent as required by linear nonequilibrium thermodynamics. However, differences
between hydraulically driven and osmotically driven transport arise at higher order
in $\X$, as explicitly calculated in (\ref{J}), (\ref{ATTRACTP}), and (\ref{REPELP}).

Continuum Poiseuille flow (zero Reynolds number solution of the Navier-Stokes equation
for flow through an infinite right circular cylinder\cite{LANDAU}) expressions (where $J
\propto r_{p}^4$) have been proposed to describe transport through microscopic channels.
Although these continuum relations have no validity in {\it microscopic} settings, they
are nevertheless often used to obtain estimates of permeabilities, pore radii
\cite{GALEY}, and membrane pore densities. These estimates typically do not agree well
with independent measurements of pore radii and pore permeabilities\cite{GALEY}. 
Although the $r_{p}$ dependences in our model are valid only for a limited range of pore
radii, they are consistent with the inapplicability, at microscopic scales, of the
continuum expression $P_{os} \propto r_{p}^4/(\eta L)$, where $\eta$ is the bulk solvent
viscosity.

Fluid flow through small orifices has been described as being both ``viscous'' and
``diffusive.'' Even though the limit (\ref{POS1}) appears to be ``viscous'' due to the
$L^{-1}$ dependence, and and (\ref{POS2}) ``diffusive,'' we see that both pressure and
osmosis driven flows in the present model is diffusive in nature.  Consider the time
averaged rate equation for occupation at site $i$,

\begin{equation}
\begin{array}{ll}
\displaystyle \dot{\sigma}_{i} & \displaystyle =
-p\sigma_{i}(1-\sigma_{i+1})-p\sigma_{i}(1-\sigma_{i-1}) \\
\: & \hspace{.7in}+p(1-\sigma_{i}) (\sigma_{i+1}+\sigma_{i-1}) \\[13pt]
\: & \displaystyle = p(\sigma_{i+1}-2\sigma_{i}+\sigma_{i-1})
\label{DIFFUSION}
\end{array}
\end{equation}

\noindent which on scales $\gg \ell$ is a diffusion equation with $p\ell^2 \cong D_{c}$ a
{\it cooperative} diffusion coefficient.  When particles interact along the chain
the overall flux is determined by the cooperative diffusion of particles.  In the simple
one-dimensional model presented, cooperative diffusion is related directly to the
precisely defined dynamical rules (\ref{RULES}), but in general is defined by linear
density-density correlation functions, or Green-Kubo relations. 

The neglect of true convective, or momentum transfer effects can be neglected in the
systems we consider. The signature of convective flow, such as Poiseuille flow, is  a
particle velocity distribution function with is Maxwellian centered about a small finite
stream velocity. However, in 1D, or nearly 1D pores, the particles thermally equilibrate
with the stationary pore walls very quickly and are unable to collectively transfer a
stream momentum.  This is particularly valid for repelling pores where entropic pressure
results in low pore occupations $\s_{i} \ll 1$.  In this large Knudsen number regime
\cite{CUSSLER}, the particles will collide with walls more often that with neighboring
particles.

\section{Extensions and Conclusions}

An extension of the 1D chain represented by (\ref{RULES}) to include a distribution of
site-dependent internal hopping rates $p_{i}=q_{i}$ can be straightforwardly calculated,
as outlined in the Appendix.  Biological realizations of internal pore defects with site
dependent hopping rates are gramicidin A channels comprising of two barrel structures
joined at the defect (membrane bilayer midplane)\cite{ALBERTS}.  Similarly, water
permeation through a bilayer membrane can be viewed as pores along the hydrophobic lipid
tails with the bilayer midplane and possible unsaturated bonds acting like defects. 
Channel proteins with widely varying internal interactions would also be modified
according to the results found in the Appendix.  Strong particle-pore interactions which
are {\it incommensurate} with $\ell$ may also lead to nonlinear transport. Here,
possible cluster diffusion \cite{SHOLL} and locked-sliding phases such as that found in
Frenkel-Kontorowa \cite{CHAIKIN} and friction models \cite{FRICTION} can also affect
transport. These effects offer the possibility of interesting nonlinearities amenable to
numerical study.  Solvents that have strong mutual attractive interactions can also be
considered. For example, a particle may hop with different rates into an empty site at
$i+1$ depending on whether or not there is an attractive particle behind it, at site
$i-1$.  The probability of such a hop to the right from $i \rightarrow i+1$ can be
represented by $p\s_{i}(1-\s_{i+1})(1-\s_{i-1}) +p'\s_{i}(1-\s_{i+1})\s_{i-1}$ where $p$
represents the rate of an isolated particle, while $p' \ll p$ represents the rate in the
presence of another particle at $i-1$ binding to and preventing the hop of the particle
at $i$ to $i+1$. The resulting particle density profile along the chain will no longer
be linear. These effects offer the possibility of interesting nonlinearities amenable to
numerical study.

For pores that are slightly larger, with radii larger than a
few times the particle diameters, similar stochastic treatments
can be applied provided particle momenta relax much faster
than particle positional degrees of freedom. Only then is the
collective stream velocity zero, and $p=q$ can be invoked. 
Near the single file limit, the pore walls are in constant
interaction with the transported molecules and randomizing
their momentum. Nevertheless, for larger pores, we can estimate
the relative rates of positional and momentum relaxation. 
Pressure (either hydrostatic or osmotic) driven flow through a
{\it macroscopic} circular pipe is described by Poiseuille's
law \cite{LANDAU}

\begin{equation}
v(r) = {\D P \over 4\eta L}(r_{p}^2-r^2)
\label{POISEUILLE}
\end{equation}

\noindent where $v(r)$ is the fluid element velocity
along the pipe, and $\eta$ is the bulk fluid viscosity.
Equation (\ref{POISEUILLE}) has often been extended to
describe flow through microscopic pipes, especially
biological membrane pores, having radii $r_{p}$ of
Angstroms. This is the origin of the notion that
osmosis through microscopic pores is ``convective.''
However, for the Poiseuille stream convection to be
comparable to ``driven diffusive transport,'' the ratio

\begin{equation}
\mbox{Pe}\equiv {vL\over D_{c}}\simeq 
{\D P r_{p}^{2} \over  4\eta D_{c}} \sim  1,
\label{PECLET}
\end{equation}

\noindent where $D_{c}$ is the effective collective
diffusion coefficient which describes dissipative particle
density relaxation, and is also approximately the momentum
diffusion, or sound attenuation coefficient. For a
hydrostatic pressure difference of $\D P = 2.5$ atm
(corresponding to a 100mM solute concentration difference),
and bulk viscosity and attenuation coefficient $\eta =
0.01$g cm/s, and $D_{c}\simeq \eta/\rho$, convective, or
``viscous''\cite{CRACKNELL} flow will be irrelevant for
$r_{p} \ll 1000\mbox{\AA}$.  For typical zeolites and
biological pores under the above conditions, $r_{p} \sim
3-8\mbox{\AA}$, $\mbox{Pe} \sim 10^{-5} \ll 1$.  Thus,
``convective'' flow described by (\ref{POISEUILLE}) is
negligible compared to diffusive transport in systems of
present concern.  Diffusive transport in this context should
not be confused with tracer diffusion or collective
diffusion in binary mixtures, rather, it is the consequence
of microscopic momentum relaxation due to frequent
collisions with the pore interior.  These collisions
increase relative to particle-particle interactions for
smaller and smaller pores, where the inner surface to pore
volume ratio increases.

Additional support for the basic assumption of diffusive
particle motions is found from  nonequilibrium molecular
dynamics studies (NEMD) of flow through a slit \cite{CRACKNELL}
where an external driving force $q\neq p$ was not used
explicitly.  However, the two reservoirs were held at different
chemical potentials, or pressures.  The density profile for
Lennard-Jones particles with attraction energies close to
$k_{B}T$, driven across 50\AA long slits of widths $2.5$ times
the particle diameters with a 35 atm hydrostatic pressure
difference was found to be linear, suggesting validity of
(\ref{SIGMA}) and Fick's law even under extreme conditions
\cite{CRACKNELL}. The validity of (\ref{PECLET}) and diffusive
($p=q$) flow suggest that pores with radii $\sim
10-50\mbox{\AA}$ a few times molecular diameters can also be
treated by multidimensional exclusion models consisting of a
series of 1D chains with particle interchange among them.
Solvent-solvent attractive interactions and diffusion of small
clusters can also be straightforwardly  implemented.  

Although we have treated only a single pore, actual experiments measure
flow over an area containing an ensemble of pores.  The averaged steady
state flux neglecting unstirred layers would then be

\begin{equation}
\langle J \rangle = \sum_{N,\{p_{i}\}}f(N,\{p_{i}\})
J(N,\{p_{i}\})
\label{AVERAGE}
\end{equation}

\noindent where $f(N,\{p_{i}\})$ is the number distribution for pores of length $N$,
with the set $\{p_{i}\}$ of internal defects. The pores in (\ref{AVERAGE}) need not be
straight, and the sum over $N \simeq d/\ell \rightarrow \infty$ represents a
distribution of pores with various total arc-lengths traversing a membrane of thickness
$d$.  We have not considered the nonlinear effects of solute-pore interactions, which we
treat in detail in the companion paper.  

The main results of this study is a thermodynamically consistent model of osmosis and
pressure driven flows through nanoscopic-sized pores. All parameters in the theory
are microscopically motivated yet are general enough to allow predictions in trends
and identify correlations between transport rates and macroscopic thermodynamic
properties. Predictions from this transport model are the site occupancies
(\ref{SIGMA}), the dependences of $\X$ on $\D \chi_{s}$ and $\D P =
P^{R}-P^{L}$, the Arrhenius temperature dependences, the pore radius $r_{p}$
dependences, and the flow maximum (\ref{MAXIMUMA}-\ref{MAXIMUMB}) as a function of
pore-solvent binding constant $\ba$. Also, microscopic expressions are derived for
phenomenologically measured permeabilities $P_{os}$ and $L_{p}$.
Equation (\ref{AVERAGE}), and the microscopically estimated values for the
parameters involved offer an reasonable, semi-quantitative description of osmosis
and pressure-driven flow through microscopic channels. The analytic model presented
provides a simple, physically motivated analysis that complements more complicated
numerical simulations by predicting rich parameter dependences.

\acknowledgements

The author thanks D. Lohse, D. Ronis, E. G. Blackman, I. Smolyarenko, T.
J. Pedley, and A. E. Hill, for discussions, and the The Wellcome Trust
for financial support.  

\begin{appendix}

\section{Flow through disordered pores}

The exact steady state flow through a pore with random internal hopping rates $q_{i}=
p_{i}$  can be found by extending the uniform pore solution (\ref{JBAR}).  When these
random hopping rates are time independent, or fluctuate independently of the particle
occupations $\s$, the steady state between section $i$ and $i+1$ is

\begin{equation}
J_{i} = p_{i}(\s_{i}-\s_{i+1})
\label{JI}
\end{equation}

\noindent and the mean-field solution becomes exact. Summing 
(\ref{JI}) across the channel yields

\begin{equation}
J\sum_{i=1}^{N-1}p_{i}^{-1} = \s_{1}-\s_{N},
\end{equation}

\noindent which upon solving with (\ref{SIGMA}) for $J$,

\begin{equation}
J = {\a\b-\g\d \over (\a+\g)(\b+\d)\sum_{i=1}^{N-1}p_{i}^{-1} + (\a+\b+\g+\d)}.
\label{JDISORDER}
\end{equation}

\noindent The above result also constitutes a steady state
solution to the XXZ spin chain Hamiltonian \cite{SANDOW} with
site-dependent energies.

\end{appendix}

\end{document}